# Constant-adiabaticity RF-pulses for generating long-lived singlet spin states in NMR

Bogdan A. Rodin,[a,b] Kirill F. Sheberstov,[a,c] Alexey S. Kiryutin,[a,b]
Joseph T. Hill-Cousins,[d] Lynda J. Brown,[d] Richard C. D. Brown,[d]
Baptiste Jamain,[a,b,e] Herbert Zimmermann,[f] Renad Z. Sagdeev,[a,b,g]
Alexandra V. Yurkovskaya,[a,b] Konstantin L. Ivanov*[a,b]

[a] *International Tomography Center, Siberian Branch of the Russian Academy of Science, Novosibirsk, 630090, Russia; Email: ivanov@tomo.nsc.ru*
[b] *Novosibirsk State University, Novosibirsk, 630090, Russia*
[c] *State Scientific Center of the Russian Federation "State Research Institute for Chemistry and Technology of Organoelement Compounds" (SSC RF GNIIChTEOS), Moscow 111123, Russia*
[d] *School of Chemistry, University of Southampton, Southampton, SO17 1BJ, UK*
[e] *Université de Toulouse, INSA Toulouse, Département de Physique, 31077 Toulouse Cedex 4, France*
[f] *Max-Planck-Institut für Medizinische Forschung, Dept. of Biomolecular Mechanisms, Heidelberg, 69028, Germany*
[g] *Kurnakov Institute of General and Inorganic Chemistry, Russian Academy of Science, Moscow, 19991, Russia*

\* Corresponding author, email: ivanov@tomo.nsc.ru, Tel.: +7(383)333-3152

**Abstract**

A method is implemented to perform "fast" adiabatic variation of the spin Hamiltonian by imposing the constant adiabaticity condition. The method is applied to improve the performance of singlet-state Nuclear Magnetic Resonance (NMR) experiments, specifically, for efficient generation and readout of the singlet spin order in coupled spin pairs by applying adiabatically ramped RF-fields. Test experiments have been performed on a specially designed molecule having two strongly coupled $^{13}$C spins and on selectively isotopically labelled glycerol having two pairs of coupled protons. Optimized RF-ramps show improved performance in comparison, for example, to linear ramps. We expect that the methods described here are useful, not only for singlet-state NMR experiments, but also for other experiments in magnetic resonance, which utilize adiabatic variation of the spin Hamiltonian.

## I. Introduction

Singlet-state Nuclear Magnetic Resonance (NMR) and Long-Lived spin States (LLSs) are emerging concepts in magnetic resonance[1-3]. Such states are protected by symmetry and relax much slower than spin magnetization. The simplest example of an LLS is given by the singlet state of a pair of spins. Singlet spin order is immune to dipolar relaxation, which is often the dominant relaxation mechanism, and can relax to equilibrium only through less efficient mechanisms[4, 5]. In the case of protons, LLS lifetimes, $T_{LLS}$, can be about 50 times longer than $T_1$-relaxation times[6]. For other nuclei, $T_{LLS}$ can be of the order of 10 minutes or even an hour[7, 8]. Such extended lifetimes of the spin order allow one to harness LLSs for probing various slow processes[9-14] and for storing spin hyperpolarization[15-23].

In spin pairs at thermal equilibrium the long-lived spin mode is not populated because there is no singlet-triplet imbalance of populations, i.e., the population of the singlet state is equal to the average population of the three triplet states. At the same time, the singlet state cannot be observed directly by NMR methods. For these reasons, experiments with singlet LLSs are usually performed in the following way[1]. In the first step, magnetization-to-singlet conversion, M2S, spin magnetization is converted into the singlet-triplet imbalance of populations and singlet spin order is generated. Then, in the second step, the singlet LLS is sustained during a variable time interval. Finally, in the third step backward S2M conversion is performed and the LLS is converted back into observable spin magnetization. The efficiency of such a M2S/S2M conversion is limited by unitary bounds[24]: the highest theoretically allowed conversion efficiency is equal to 2/3. In this work, we aim to approach this limit of the conversion efficiency. To this end, we exploit a recently proposed method termed APSOC (Adiabatic-Passage Spin Order Conversion)[25-28], which is expected to provide the desired efficiency of 2/3. Previous studies using APSOC provided a lower M2S/S2M conversion efficiency,[25, 27] presumably, because of spin relaxation during M2S/S2M conversion leading to a loss of the spin order. Generally, adiabatic processes imply a slow variation of some externally controlled parameter of the spin Hamiltonian; hence, they require a considerably long time with a consequence that relaxation effects become detrimental leading to a loss of spin order. To tackle this problem we worked with a molecule having extremely long LLS lifetime[7] and also optimized the conversion stage. In addition, to demonstrate the versatility of APSOC we applied APSOC to the glycerol molecule with two pairs of weakly-coupled protons having a long-lived state in the presence of a spin-locking filed. The optimization method is streamlined to provide "fast" adiabatic variation. To perform such a variation of the RF-ramps we set the time dependence of the RF-field amplitude such that the adiabaticity parameter is constant in time[29-33]. In addition to APSOC, we apply this method to improve the performance of the SLIC (Spin-Locking Induced Crossing) method[34] in its variant[35] with adiabatically ramped RF-field. As we demonstrate below, this strategy allows us to get close to the desired conversion efficiency. Furthermore, the optimized APSOC method is efficient for both weakly-coupled and strongly-coupled systems.

## II. Methods
### A. APSOC and SLIC techniques

All experiments were conducted using a 700 MHz NMR spectrometer with the $B_0$ field of 16.4 T. APSOC experiments were performed using the protocol shown in **Figure 1** (top). The protocol comprises 4 stages:

Stage 1: M2S conversion performed by the RF$_1$-field, which is ramped up in an adiabatic way. As has been shown previously[25-28], such an RF-field can transfer the population of the triplet $|T_+\rangle$ (or $|T_-\rangle$) spin state to the singlet state, $|S\rangle$. Since at equilibrium conditions the $|T_+\rangle$ state is overpopulated (for protons or any other nuclei with positive gyromagnetic ratio) and the $|T_-\rangle$ state is under populated (spin magnetization is given by their population difference) this stage is essentially M2S conversion. To do so, one needs to set the RF-frequency, $\nu_{rf}$, precisely and to make sure that the RF-ramp is compatible with adiabatic variation of the spin Hamiltonian. As far as the choice of $\nu_{rf}$ is concerned, the off-set of $\nu_{rf}$ from the "center of the spectrum", $\langle v \rangle$, of the spin pair has to be non-zero, but should not exceed the half-width of the spectrum[25-28]. Interestingly, by varying the sign of the off-set, $(v_{rf} - \langle v \rangle)$, one can choose the conversion pathway, which is either $T_+ \to S$ or $T_- \to S$. The maximal strength of the RF$_1$-field, $v_1^{max}$, should be set such that $|S\rangle$ is an eigen-state of the system. The duration of the RF$_1$-ramp, $\tau_{sw}$, and time profile, $v_1(t)$, should be set such that the switch is adiabatic. Optimization of the $v_1(t)$ profile is described below.

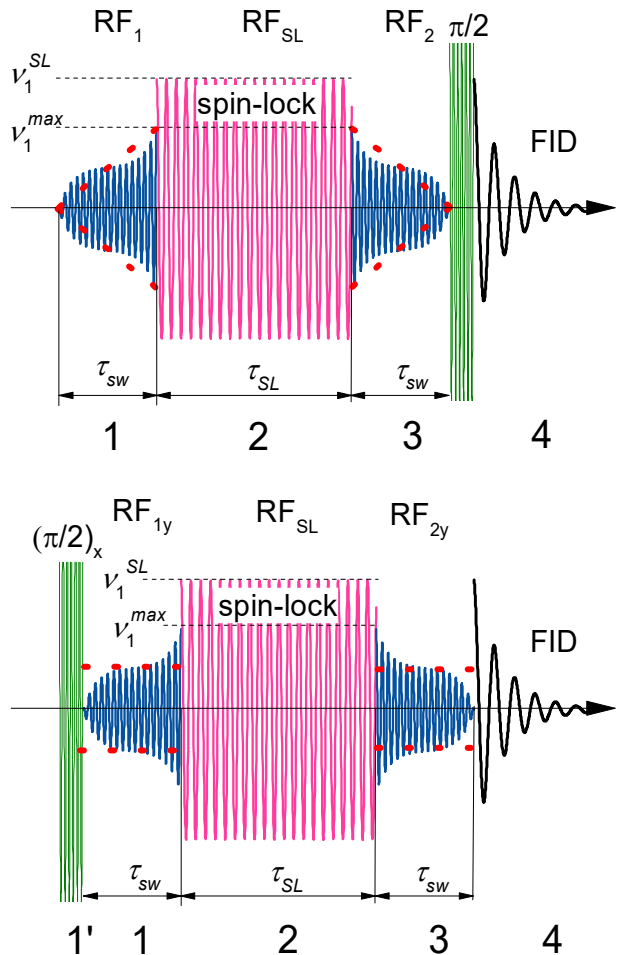

**Figure 1.** (top) Protocol of the APSOC experiment comprising 4 stages. In stage 1, M2S conversion is performed by an adiabatically ramped RF-field denoted as RF$_1$. In stage 2 the LLS is sustained (for a strongly coupled spin system there is no need to apply a spin-locking RF-field at this stage). In stage 3, S2M conversion, the RF$_2$-field is applied, which is adiabatically reduced to zero. Finally, in stage 4 an NMR pulse is applied and the NMR spectrum is obtained as the Fourier transform of the FID signal. See text for further explanation of APSOC and the filtering method. (bottom) Protocol of the SLIC experiment. Stages 1-3 are the same as for the APSOC sequence. In stage 1' longitudinal thermal magnetization is converted to the transverse magnetization by a hard $\frac{\pi}{2}$ pulse. In stage 4 the NMR spectrum is obtained as the Fourier transform of the FID signal (without using additional pulses).

Stage 2: LLS maintenance during the time period $\tau_{SL}$. This stage is performed in a different way depending on the parameters of the spin pair, namely, on the relation between the J-coupling, $J$, and the difference in NMR frequencies, $\delta v$. When the spin pair is weakly coupled, $|\delta v| \gg |J|$, applying a spin-locking field is required for sustaining the LLS. However, in the opposite case of a strongly coupled spin system, $|J| \gg |\delta v|$, the LLS is sustained even in the absence of an additional RF-field.

Stage 3: S2M conversion is performed by the RF$_2$-field, which is ramped down in an adiabatic way. Since adiabatic transitions are reversible, the RF$_2$-field can be obtained from the RF$_1$-field by simply reverting it. This RF-ramp provides the $S \to T_+$ or $S \to T_-$ conversion. As a result, magnetization is formed from the singlet order.

Stage 4: detection of the resulting longitudinal magnetization by applying an NMR pulse, recording the Free Induction Decay (FID) signal and performing Fourier transformation of the FID.

We also emphasize two more beneficial features of using APSOC[25]. First, the technique is applicable to any spin pair, i.e., it provides the desired conversion for arbitrary relation between $J$ and $\delta\nu$. Second, the method can be modified to suppress all unwanted signals by arranging a pseudo phase cycle. To do so, the experiment should be repeated at least twice at different frequencies of the RF$_1$ and RF$_2$-fields in order to switch between the $T_+ \leftrightarrow S$ and $T_- \leftrightarrow S$ conversion pathways. Such a method gives rise to the SOS-filter (SOS=Singlet Order Selection)[26].

In addition to APSOC we exploited a variant of SLIC[35], which is applicable to strongly coupled spin pairs, **Figure 1** (bottom). In this case for the M2S conversion we applied a hard $\frac{\pi}{2}$-pulse generating transverse magnetization (from the initial longitudinal magnetization) followed by an additional RF-pulse. This RF-pulse is phase-shifted by 90 degrees with respect to the first pulse and its amplitude is adiabatically ramped from zero to $\nu_1^{max}$. As has been shown before, such an RF-pulse converts the transverse magnetization into the long-lived singlet order. LLS maintenance is performed in the same way as in APSOC (a spin-locking RF-field is applied); for the S2M conversion we ramp down the RF-field from $\nu_1^{max}$ to zero in an adiabatic fashion. After that the FID signal arising from the resulting transverse magnetization is detected and the NMR spectrum is obtained. In the conventional variant of the SLIC technique the $\nu_1$ value is set equal to $|J|$ and spin order conversion is due to excitation of spin coherences. As we demonstrate here, this method provides a lower conversion efficiency, as it requires precise setting of experimental parameters.

For measuring $T_{LLS}$ one should repeat the experiment with variable $\tau_{SL}$ values. The resulting $\tau_{SL}$-dependence of the resulting signal intensity, $I(\tau_{SL})$, should be modeled by the following function:

$$I(\tau_{SL}) = I_0 + A_{fast} \exp\left[-\frac{\tau_{SL}}{T_{fast}}\right] + A_{LLS} \exp\left[-\frac{\tau_{SL}}{T_{LLS}}\right] \qquad (1)$$

Here the fast component stands for relaxation within the triplet states whereas the slow component is due to the LLS, $T_{LLS} \gg T_{fast} \sim T_1$. The coefficient, $A_{LLS}$, gives the weight of the long-lived component: this coefficient is equal to 2/3 when the M2S/S2M conversion is perfectly optimized and there is no loss of spin order during the RF-ramps. Hence, here we aim to achieve $A_{LLS} = 2/3$.

### B. Optimization of the conversion stage

As explained above, RF-ramps need to be adiabatic. Literally, adiabatic variation of the Hamiltonian means that $\nu_1(t)$ should be varied slowly. However, when the $\tau_{sw}$ time is of the order of $T_1$ and $T_2$-relaxation times, spin order is irreversibly destroyed during the RF-switch. Hence, for optimization we propose minimizing the $\tau_{sw}$ time still keeping high degree of adiabaticity. In the present case the Hamiltonian of a pair of spins ½, $a$ and $b$, in the RF-rotating frame is of the form (in $h$ units):

$$\widehat{H}(t) = -(\nu_a - \nu_{rf})\hat{I}_{az} - (\nu_b - \nu_{rf})\hat{I}_{bz} + J(\hat{\mathbf{I}}_a \cdot \hat{\mathbf{I}}_b) - \nu_1(t)[\hat{I}_{ax} + \hat{I}_{bx}] \qquad (2)$$

When $\widehat{H}(t)$ changes with time in an adiabatic way (due to the time-dependent RF-field strength, $\nu_1(t)$) populations follow instantaneous eigen-states. Generally, for a pair of time-dependent adiabatic states $|i\rangle$ and $|j\rangle$ the adiabaticity parameter is as follows:

$$\xi_{ij} = \frac{\left\langle i \left| \frac{d}{dt} \right| j \right\rangle}{\omega_{ij}} \qquad (3)$$

Here the numerator gives the rate at which the eigen-states change with time, while the denominator gives the internal evolution frequency of the system, $\omega_{ij} = 2\pi|E_i - E_j|$ ($E_i$ is the eigen-value of $\widehat{H}$ corresponding to the $|i\rangle$ eigen-state). In practice, it is more convenient to calculate $\xi_{ij}(t)$ using the formula (mathematically equivalent to eq. (3)):

$$\xi_{ij}(t) = \frac{\left\langle i \left| \frac{d\widehat{H}}{dt} \right| j \right\rangle}{\omega_{ij}^2} \tag{4}$$

In practice, it is easier to compute the time derivative of $\widehat{H}(t)$, which is simply equal to

$$\frac{d\widehat{H}(t)}{dt} = -[\hat{I}_{ax} + \hat{I}_{bx}]\frac{dv_1(t)}{dt} = -\hat{I}_x \frac{dv_1(t)}{dt} \tag{5}$$

then $d|j\rangle/dt$. When $\xi_{ij} \ll 1$ variation of $\widehat{H}$ is adiabatic. In reality, both numerator and denominator in eq. (3) are relatively complex functions of time. For this reason, $\xi_{ij}$ can strongly vary with time. For instance, when $v_1(t)$ is varied at a constant speed, $\xi_{ij}$ can become large when energy levels closely approach each other, i.e., at Level Anti-Crossings (LACs), because $d|j\rangle/dt$ becomes large while $\omega_{ij}$ becomes small. To tackle this problem we propose to set the $v_1(t)$ such that $\xi_{ij}$ is constant or nearly constant during the entire variation of $\widehat{H}(t)$, i.e., we propose to use an optimized constant adiabaticity profile, $v_1(t)$. For a two-level system, there is a single $\xi_{ij} = \xi_{12}$ parameter, which should be set constant. In a multi-level system, we can introduce the generalized adiabaticity parameter, which is as follows:

$$\langle \xi \rangle(t) = \sqrt{\sum_{ij} \xi_{ij}^2(t)} \tag{6}$$

The next step is to set $v_1(t)$ such that $\langle \xi \rangle = \xi_0 = $ const during the entire variation of the RF-field strength. By using eqs. (5) and (6) we obtain the following expression:

$$\left|\frac{dv_1}{dt}\right| \sqrt{\sum_{ij} \frac{\left(\langle i|\hat{I}_x|j\rangle\right)^2}{\omega_{ij}^4}} = \xi_0 \Rightarrow \frac{dv_1}{dt} = \pm \xi_0 \left\{\sum_{ij} \frac{\left(\langle i|\hat{I}_x|j\rangle\right)^2}{\omega_{ij}^4}\right\}^{-1/2} \tag{7}$$

This expression immediately yields the time derivative $dv_1/dt$, which is consistent with $\langle \xi \rangle = \xi_0 = $ const, i.e., it provides the sought constant-adiabaticity $v_1(t)$ function. The $\xi_0$ value is unambiguously related to the total time, $\tau_{sw}$, of the RF-field switch. The RF-up profiles can be obtained from eq. (6) by taking positive $dv_1/dt$ and assuming $v_1(t = 0) = 0$; the RF-down profile can be then obtained by reverting the RF-up profile.

Experiments presented here were done using two different molecules, containing strongly and weakly coupled spin pairs.

The first molecule is a specially designed naphthalene derivative, hereafter, naphthalene* (see the structure in **Figure 2**), which has a strongly coupled spin pair comprising two nearly equivalent $^{13}$C-nuclei with a very long singlet lifetime of approximately 1 hour (at the external magnetic field strength of about 1 Tesla)[7]. In this spin pair, the J-coupling is significantly greater than the difference in NMR frequencies of the two $^{13}$C nuclei: specifically, $J = 54.6$ Hz and $\delta v = |v_a - v_b| = 10$ Hz at the $B_0$ field of 16.4 T. The spectrum of the molecule studied here is also shown in **Figure 2**: it comprises two intense central lines with a very small splitting of about $\delta v^2/2J \approx 0.9$ Hz and two low-intensity satellites, which are separated from the center of the spectrum by approximately $\pm J$. At the 16.4 T field the $T_{LLS}$ relaxation time is about 217 s, which is approximately $31 \cdot T_1$. For optimization of the RF-ramps here we used the following parameters: $v_1^{max} = 100$ Hz, $\Delta = v_{rf} - \langle v \rangle = 10$ Hz. The shape of the ramps was determined by imposing the constant-adiabaticity condition as described above.

The second experimental system we have studied is the glycerol molecule with a D-isotope label, hereafter glycerol-2-d$_1$, see the structure in **Figure 3**. Due to proton exchange, in D$_2$O this molecule is described by the formula ODCH$_2$-CD(OD)-CH$_2$OD. Hence, in the molecule there are two pairs of coupled protons, which have very small couplings to each other and thus can be treated as two non-interacting pairs of spins with identical NMR parameters, $J$ and $\delta\nu$. At $B_0 = 16.4$ T the NMR parameters are: $J = 11.7$ Hz and $\delta\nu = 62$ Hz. The spin pairs of the CH$_2$-groups also have an LLS with a lifetime of 17 s, which is approximately $7 \cdot T_1$. Hence, $\delta\nu$ is significantly larger than $J$ and the system can be treated as weakly-coupled. Nonetheless, the NMR spectrum, see **Figure 3**, exhibits the "roof-effect": the two outer NMR lines have a lower intensity as compared to the two central lines. Strictly speaking, as $\delta\nu \approx 5 \cdot J$, glycerol-2-d$_1$ is an intermediate case between strong coupling and the limiting case of weak coupling, $|J| \ll |\delta\nu|$. Generating singlet LLSs in such a situation is challenging (most methods are designed to deal with either strongly coupled or weakly coupled spin pairs); for this reason, we find it interesting to investigate the performance of APSOC in the case of glycerol-2-d$_1$.

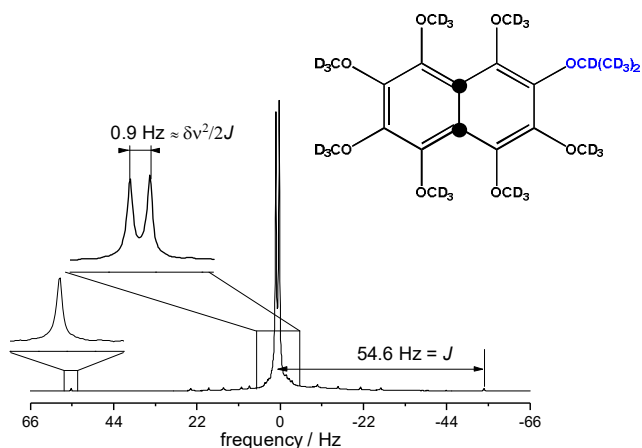

**Figure 2**. Doubly $^{13}$C-labeled naphthalene derivative and its $^{13}$C-NMR spectrum at $B_0 = 16.4$ T (the spectrum is centered at the $\langle\nu\rangle$ frequency). This molecule has $T_1 = 7$ s and $T_{LLS} = 217$ s at $B_0 = 16.4$ T in the absence of spin-locking; the $^{13}$C labels are highlighted.

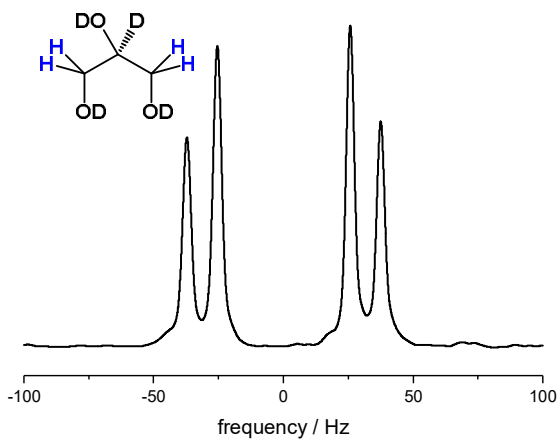

**Figure 3**. Glycerol-2-d$_1$ molecule and its $^1$H-NMR spectrum in D$_2$O at $B_0 = 16.4$ T. This molecule has $T_1 = 2.32$ s and $T_S = 17$ s at $B_0 = 16.4$ T in the presence of 1000 Hz spin-locking.

**Figures 4** and **5** describe the idea behind the APSOC and adiabatic SLIC methods in the case of a strongly-coupled spin pair, explaining how correlation of states in the rotating frame works.

In the APSOC experiment, upon increase of $\nu_1$ the $|T_+\rangle$ state is correlated with the $|S\rangle$ state. Consequently, population of the $T_+$-state flows into the $|S\rangle$ state and M2S conversion occurs. Since adiabatic transitions are reversible, upon decrease of $\nu_1$ backward S2M conversion will occur. In an APSOC experiment it is important that $\Delta \neq 0$ (otherwise the $T_+$ and $T_-$ are degenerate) and that $|\Delta| < |J|$ (otherwise correlation changes such that the singlet and triplet states never get mixed). In APSOC, it is important that the passage is adiabatic, in particular, at the LAC region at $\nu_1 \approx |J|$. An example of optimization of RF-ramps for APSOC is shown in **Figure 4**. One can see that the optimization suggests slow variation of $\nu_1(t)$ in two regions of parameters where energy levels closely approach each other and have LACs. One such LAC is occurring in the triplet manifold at $\nu_1 \to 0$ and the other one is occurring when the singlet level crosses with one of the triplet levels at $\nu_1 \approx |J|$.[34] Away from LACs, optimization suggests fast variation of $\nu_1(t)$.

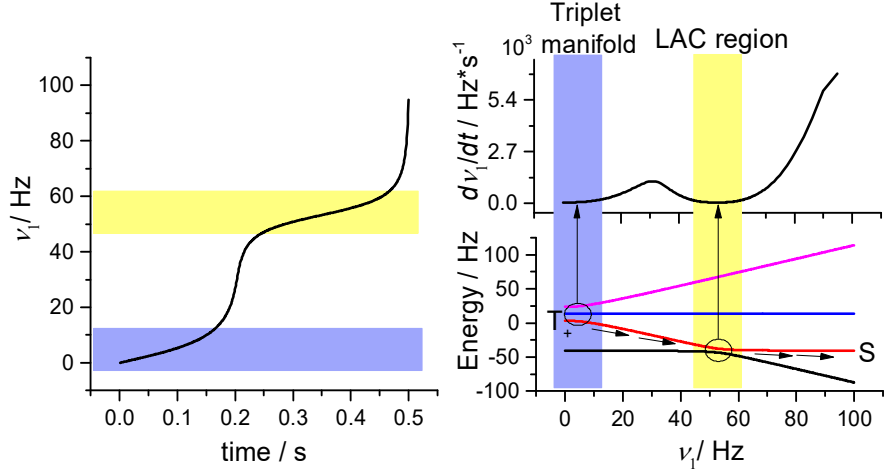

**Figure 4**. (*left*) RF-field profile for a constant-adiabaticity pulse, as optimized for the APSOC method. We highlight in blue the region where there is a LAC in triplet manifold and in yellow the singlet-triplet LAC region. (*right, bottom*) Energy levels for a strongly coupled spin pair. (*right, top*) The speed of variation of the RF-field amplitude, $d\nu_1/dt$. The parameters used in calculation are $J = 54$ Hz, $\delta\nu = 10$ Hz, $\nu_1^{max} = 100$ Hz, $\Delta = 10$ Hz, $\tau_{sw} = 0.5$ s.

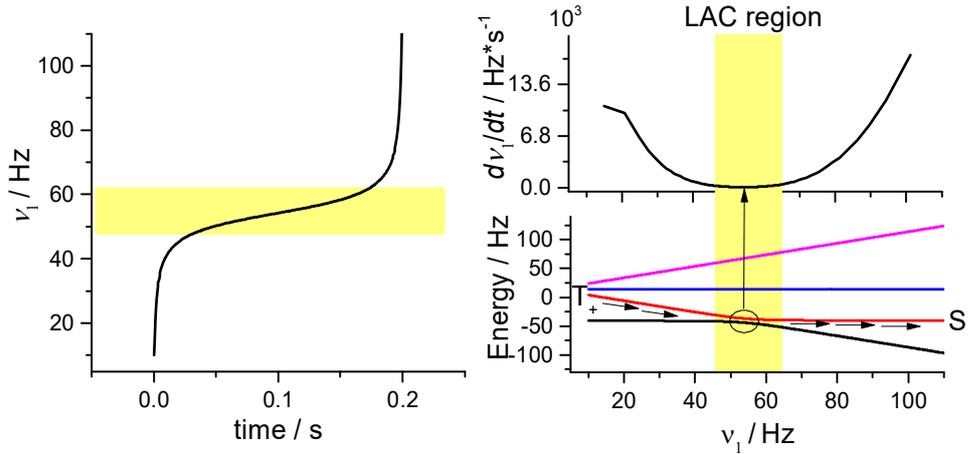

**Figure 5**. (*left*) RF-field profile for a constant-adiabaticity pulse, as optimized for the SLIC method. The yellow region corresponds to the singlet-triplet LAC region at $|\nu_1| = |J|$. (*right, bottom*) Energy levels for a strongly coupled spin pair. (*right, top*) The speed of variation of the RF-field amplitude, $d\nu_1/dt$. The parameters used in calculations are $J = 54$ Hz, $\delta\nu = 10$ Hz, $\nu_1^{max} = 100$ Hz, $\Delta = 10$ Hz, $\tau_{sw} = 0.5$ s.

Adiabatic SLIC works in a similar way by using adiabatic passage through the LAC at $\nu_1 \approx |J|$; a subtle difference between the two techniques is that SLIC works best when $\Delta = 0$. Hence, optimization can be carried out in a similar way for the adiabatically swept SLIC pulses, see **Figure 5**, suggesting slow variation of $\nu_1(t)$ at the singlet-triplet LAC at $\nu_1 \approx J$. Here we compare the performance of M2S/S2M experiments with optimized ramped RF-fields and with linear ramps.

A similar optimization has been carried out for glycerol-2-$d_1$, see discussion below. In this case we have performed optimization only for APSOC because SLIC is not designed for weakly coupled spin systems. The optimization procedure suggests that the range of $\nu_1$ values where the spins are weakly coupled is passed slowly; however, once the strong coupling regime is reached the speed $d\nu_1(t)/dt$ is increased. The resulting $\nu_1(t)$ is thus drastically different from the simple linear $\nu_1(t)$ profile, see below.

## C. Sample preparation

To achieve the maximum M2S/S2M conversion, we used a liquid sample, a 20 mM solution of naphthalene* in acetone, was placed in a small sealed insert that was put into a standard 5 mm NMR tube and fixed about 1 cm above the bottom of the tube. The NMR tube (with the insert inside) was filled with protonated acetone and sealed. In this way, we are able to place the entire sample inside the active volume of the NMR coil. This provides high $B_0$ and $B_1$ homogeneity required for our experiments. With this sample preparation method, we also do not suffer from $B_0$ drifts caused by volatility of acetone. Likewise, we minimize detrimental effects[36] of convection and diffusion on the measured $T_{LLS}$ value. Nevertheless, the diminished sample size leads to an increase of $B_1$ inhomogeneity. To minimize this unwanted effect, we placed our small sample volume into the larger NMR tube filled with the same solvent.

In experiments with glycerol-2-$d_1$ we used a 50 mM $D_2O$ (Deutero GmbH, Germany) solution in a standard 5 mm NMR tube. We added ethylendiaminetetraacetic acid at 40 mM concentration and bubbled the sample with nitrogen for 10 minutes to remove paramagnetic agents that facilitate spin relaxation. pH* of the solution was 11.5. To synthesize selectively labelled glycerol we used the following procedure. Commercial diethylmalonate was oxidized to diethyl-acetoxymalonate with lead tetraacetate, according to Ref. 37. The 2-acetoxy-2-H-diethylmalonate was exchanged to the corresponding 2-acetoxy-2-D-diester at room temperature with $D_2O$ and potassium carbonate as the base within 4 hours, extracted 4 times with diethyl-ester and distilled. This deuteration procedure was repeated twice. NMR proved a deuteration degree 2-D better than 98%. The deuterated 2-D-2-acetoxydiethylmalonate was reduced under reflux for one hour to the 2-D-glycerol with $LiAlH_4$ suspended in diethyl-ester. After hydrolysis with 5% $H_2SO_4$, the phases were separated, the aqueous phase was adjusted to pH 7 and treated with a mixed ion-exchange resin column. Water was evaporated under reduced pressure and 2-D-glycerol was distilled with a membrane vacuum pump at around 130°C. NMR has demonstrated deuteration at the 2-D position of about 98%.

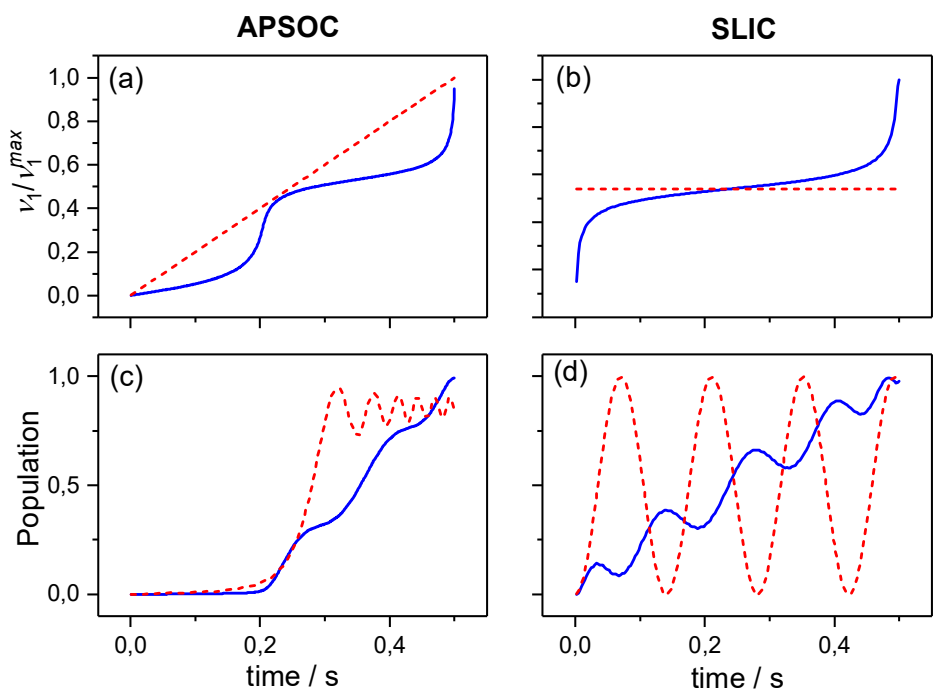

**Figure 6.** (top, a and b) Time-dependence $v_1(t)$ for optimized (blue line) and linear (red dashed line) RF ramps; optimization is performed for APSOC (a) and SLIC (b) methods. (bottom) Evolution of the singlet spin order while applying the corresponding RF-ramps for APSOC (c) and SLIC (d). In the SLIC case we compare adiabatic SLIC with the conventional SLIC experiment with excitation of the singlet-triplet spin coherence. The initial state for both simulation is $|T_+\rangle$; in the APSOC case the spins are polarized along the $z$-axis; in the SLIC case they are polarized along the $x$-axis. Simulation parameters are the same as for **Figures 4** and **5**.

## III. Results and discussion
### A. Spin order conversion in naphthalene*

Before presenting experimental results, we performed numerical simulations of M2S/S2M conversion using optimized and linear RF-ramps. These results are presented in **Figure 6** for the APSOC and SLIC techniques, respectively. We also provide a calculation of the M2S efficiency achieved with optimized and linear RF-ramps in APSOC, see **Figure 6**. One can see that optimization of APSOC provides complete M2S conversion. In the case of linear RF-ramp the conversion efficiency is also high, but not maximal. Furthermore, upon passage through the LAC at $\nu_1 \approx J$ spin coherences are excited, giving rise to oscillatory contributions to spin evolution.

For the chosen set of parameters, optimized APSOC and adiabatic SLIC have almost the same performance, which guarantees almost maximal M2S/S2M conversion efficiency. Here we also compared SLIC with an optimized RF-ramp with the conventional SLIC, which exploits a constant-amplitude RF-pulse with $\nu_1$ set equal to $J$, see **Figure 6**. In the latter case conversion is due to excitation of the singlet-triplet spin coherence. In this situation, conversion is faster and complete conversion is achieved for the pulse duration of $1/\sqrt{2}\Delta$. Despite this advantage of SLIC, here we show that in practice the methods using adiabatic switches work better, presumably, due to their greater robustness, e.g., to inhomogeneity of the $B_0$ and $B_1$ fields.

Using optimized and linear RF-ramps, we performed experiments on the $^{13}$C singlet-state in naphthalene*. The relaxation curves obtained according to the experimental protocols of APSOC and adiabatic SLIC are shown in **Figure 7**. One can clearly see that the optimized adiabatic RF-profiles exhibit a much better performance. Fitting parameters obtained using eq. (1) are shown is **Table 1**. In the fitting procedure, we used common values of $T_{fast}$ and $T_{LLS}$ for all relaxation traces; fitting gives the following relaxation times: $T_{fast} = 6.5 \pm 3$ s and $T_{LLS} = 217 \pm 8$ s. Optimization of both adiabatic SLIC and APSOC provides the same conversion efficiency giving rise to $A_{LLS} \approx 0.6$, which is close to the theoretical maximum of $A_{LLS}^{max} = 2/3$. The efficiency for single M2S (or S2M) conversion can be estimated as $\sqrt{A_{LLS}/A_{LLS}^{max}} \approx 0.95$. For APSOC with a linear RF-ramp we obtained $A_{LLS} \approx 0.35$, which is significantly lower. For the conventional SLIC method based on excitation of the singlet-triplet coherence we obtained $A_{LLS} \approx 0.38$, which is also much lower than in the case of optimized RF-ramps.

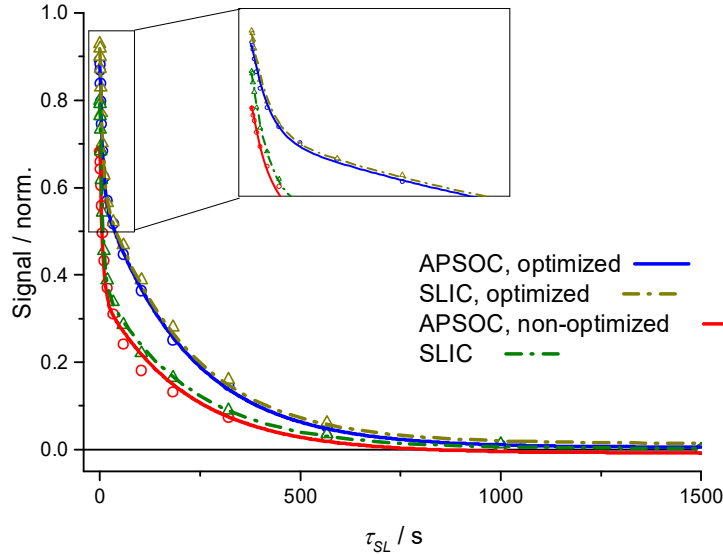

**Figure 7**. Relaxation traces obtained in APSOC and SLIC experiments on naphthalene*; here the dependence on the spin-locking time, $\tau_{SL}$ is shown. Here in APSOC we used non-optimized (linear) and optimized ramps; in SLIC experiments we used SLIC with an optimized RF-ramp and the conventional SLIC method using an RF-pulse with $\nu_1 = J$ and duration of $\frac{1}{\sqrt{2}\Delta} = 0.06$ s; the time traces are fitted by biexponential curves. The system parameters are $J = 54$ Hz, $\delta\nu = 10$ Hz. The experimental parameters are $\nu_1^{max} = 100$ Hz, $\Delta = 10$ Hz, $\tau_{sw} = 0.5$ s for both APSOC and SLIC methods.

**Table 1.** Parameters obtained from fitting of the SLIC and APSOC experiments for naphthalene*.

| Method | $I_0$ | $A_{fast}$ | $T_{fast}$, s | $A_{LLS}$ | $T_{LLS}$, s |
|---|---|---|---|---|---|
| APSOC, optimized | 0.005±0.007 | 0.30±0.01 | 6.5±0.3 | 0.59±0.01 | 217±8 |
| APSOC, linear | -0.01±0.01 | 0.34±0.01 | 6.5±0.3 | 0.36±0.01 | 217±8 |
| SLIC, optimized | 0.002±0.003 | 0.33±0.01 | 6.5±0.3 | 0.59±0.01 | 217±8 |
| SLIC | 0.03±0.07 | 0.42±0.01 | 6.5±0.3 | 0.38±0.01 | 217±8 |

We attribute lower efficiency of APSOC with linear RF-ramps to non-optimal passage through LACs, which leads to a loss of spin order. As far as the conventional SLIC technique is concerned, the loss of the M2S/S2M conversion efficiency is most likely due to $B_0$ and $B_1$ inhomogeneity.

To make sure that optimization proposed here is robust to fluctuations of parameters and possible inaccuracies in setting parameters of RF-fields we have run experiments and calculations with systematic variation of parameters such as $\Delta$, $\nu_1^{max}$ and $\tau_{sw}$. The results are shown for APSOC and SLIC in **Figures 8** and **9**, respectively. In both cases the optimized RF-ramps are robust to variation of the experimental parameters: in the dependences of the conversion efficiency on each parameter there is a flat maximum, which is sufficiently broad for our purposes. The experimental results are well reproduced by numerical simulations. The discrepancy between the experiment and theory at long switching times $\tau_{sw}$ is most likely due to relaxation processes, which are not taken into account in simulations. This study shows that for practical purposes the optimization proposed here is reasonably stable to errors in setting experimental parameters.

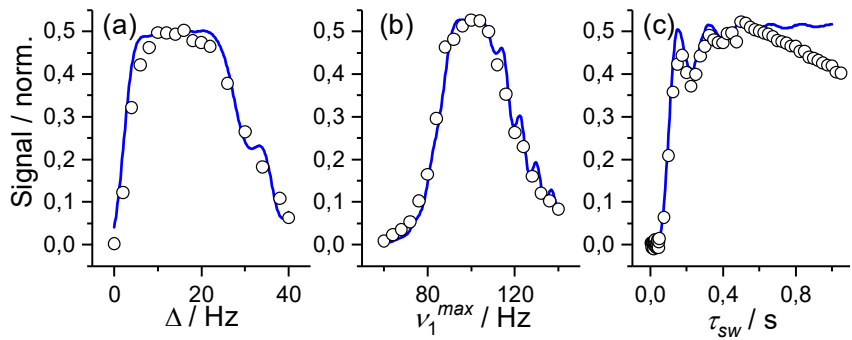

**Figure 8.** (top) Performance of the APSOC method upon variation of the RF-frequency $\Delta$ (a), its maximal amplitude $\nu_1^{max}$ (b), and switching time $\tau_{sw}$ (c). The optimal experimental parameters are $\Delta = 10$ Hz, $\nu_1^{max} = 100$ Hz, $\tau_{sw} = 0.5$ s. All signals are normalized to the thermal NMR signal intensity.

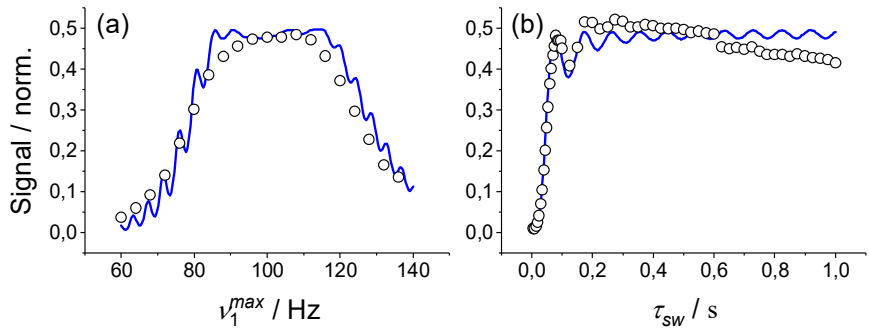

**Figure 9.** Performance of the adiabatic SLIC method upon variation of the RF-field amplitude $\nu_1^{max}$ (A), and the switching time $\tau_{sw}$ (B). All signals are normalized to the thermal NMR signal intensity.

To complete the discussion of robustness of different methods, in this subsection we present the theoretical dependence of the conversion efficiency on the inhomogeneity of the $B_0$ and $B_1$ fields for the conventional SLIC and for its version using an adiabatic $\nu_1$ sweep. The analysis of such a dependence is of practical importance allowing one to estimate the factors limiting the efficiency of SLIC: while the theoretical efficiency of SLIC should be as high as that for APSOC and adiabatic SLIC, in experiments it is much lower. Our analysis shows, see **Figure 10**, that the conventional SLIC method is very sensitive the inaccuracies in setting the $\nu_1$ value: once $\nu_1$ is such that the matching condition is no longer fulfilled the conversion efficiency drops. Hence, once the $B_1$-field is inhomogeneous over the sample volume, for some positions in the sample low conversion efficiency is expected and $A_{LLS}$ is supposed to decrease. At the same time, SLIC is reasonably tolerant to inaccuracies in setting the $\Delta$ value, which may be coming, e.g., from the $B_0$ inhomogeneity. As far as SLIC with adiabatic $\nu_1$ variation is concerned, it is reasonably tolerant to the variation of $\Delta$ being much less sensitive to the $B_1$ inhomogeneity (i.e., to the $\nu_1^{max}$ variation). This analysis, which is consistent with previous results of Theis et al.[35], gives a possible reason why the SLIC method is inefficient for the sample under study. Hence, methods utilizing adiabatic correlation of states upon variation of parameters of the spin Hamiltonian are advantageous as they are less sensitive to small inaccuracies in setting experimental parameters.

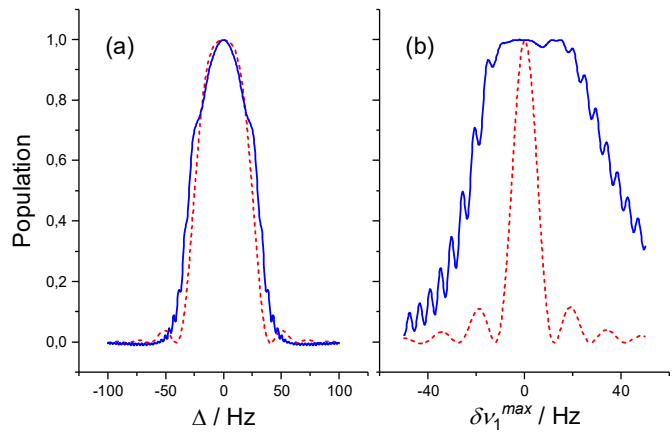

**Figure 10**. The performance of adiabatic (blue line) and conventional SLIC (red dashed line) upon variation of RF frequency $\Delta$ from spectra center (a) and RF maximum amplitude $\delta\nu_1^{max}$ (b).

### B. Spin order conversion in glycerol-2-d$_1$

We have also performed a similar study for glycerol-2-d$_1$ in D$_2$O. We have compared ASOC with linear RF-ramp with APSOC using the optimized "constant-adiabaticity" RF-ramp. The two RF-ramps are shown in **Figure 11**, along with the calculated conversion efficiencies. One can see that optimization suggests a slow $\nu_1(t)$ variation as small $\nu_1$ values followed by a faster $\nu_1(t)$ variation. Hence, the RF-field profile strongly differs from linear variation. One can also see that optimization strongly increases the conversion efficiency.

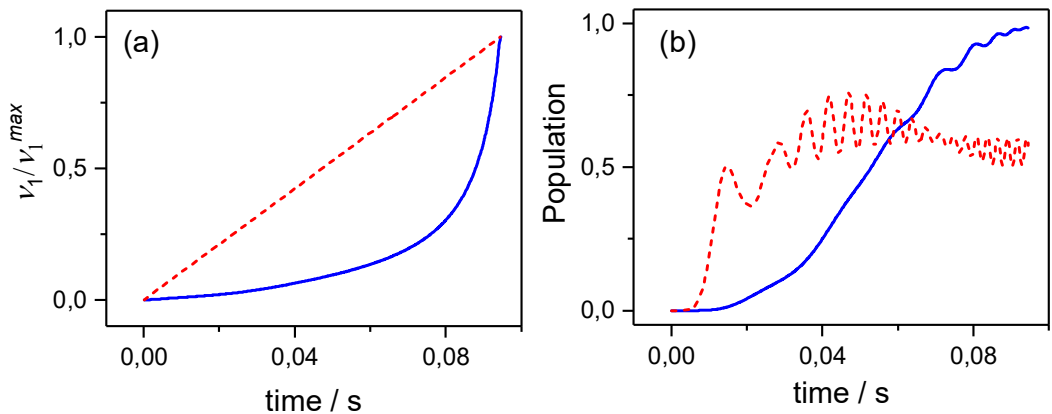

**Figure 11**. (left) Time-dependence $\nu_1(t)$ for optimized (blue line) and linear (red dashed line) RF ramps; optimization is performed for the APSOC method. (right) Evolution of the singlet spin order while applying the corresponding RF-ramps. The initial state for both simulation is $|T_+\rangle$; the spins are polarized along the $z$-axis. Simulation parameters are $J = 11.7$ Hz, $\delta\nu = 62$ Hz.

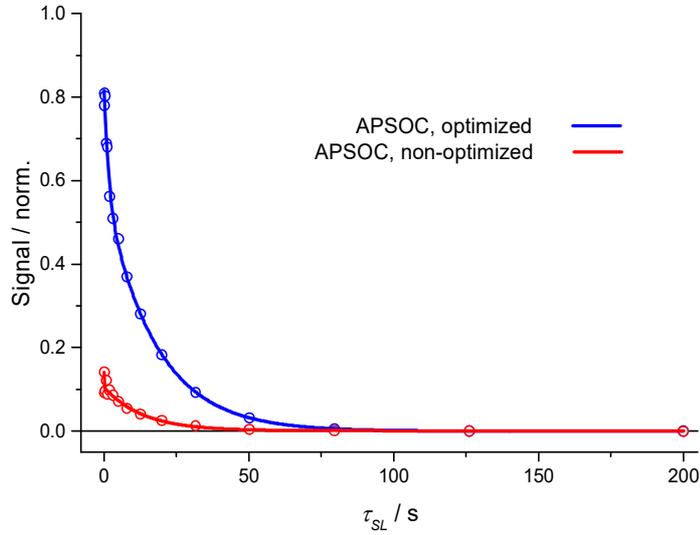

**Figure 12.** Relaxation traces obtained in APSOC experiments on glycerol-2-d$_1$ in D$_2$O; here the dependence on the spin-locking time, $\tau_{SL}$ is shown. Here in APSOC we used non-optimized (linear) and optimized ramps. The system parameters are $J = 11.7$ Hz, $\delta\nu = 62$ Hz. The experimental parameters are $\nu_1^{max} = 410$ Hz, $\Delta = 10$ Hz, $\tau_{sw} = 95$ ms.

Experiments clearly demonstrate that optimization dramatically increases the performance of APSOC, see **Figure 12**. We have run the APSOC experiments for both kinds of RF-ramps and varied $\tau_{SL}$ to investigate the LLS. The experimental $\tau_{SL}$ dependences are bi-exponential with $A_{fast} \approx 1$ s and $A_{LLS} \approx 15$ s. The values of $A_{LLS}$ are remarkably different for the non-optimized APSOC and for optimized APSOC, being equal to 0.1 and 0.58, respectively. Hence, optimization allows us to achieve the conversion efficiency that is close to the theoretically allowed maximum. The fitting parameters are summarized in **Table 2**.

**Table 2.** Parameters obtained from fitting the APSOC experiments for glycerol-2-d$_1$.

| Method | $I_0$ | $A_{fast}$ | $T_{fast}$, s | $A_{LLS}$ | $T_{LLS}$, s |
|---|---|---|---|---|---|
| APSOC, optimized | 0.005±0.007 | 0.27±0.04 | 1.4±0.5 | 0.59±0.04 | 17±2 |
| APSOC, linear | 0.001±0.007 | 0.15±0.3 | 0.5±0.1 | 0.10±0.01 | 15±3 |

Like in the case of naphtalene*, we have analyzed the robustness of the proposed APSOC optimization. The results of this analysis are presented in **Figure 13**. One can see that optimization provides a sufficient stability with respect to variation of the experimental parameters ($\Delta$, $\nu_1^{max}$ and $\tau_{sw}$): all dependences have maximum, which is reasonably broad for running APSOC experiments in an efficient way. The experimental dependences on $\Delta$, $\nu_1^{max}$ and $\tau_{sw}$ are in good agreement with the simulation results.

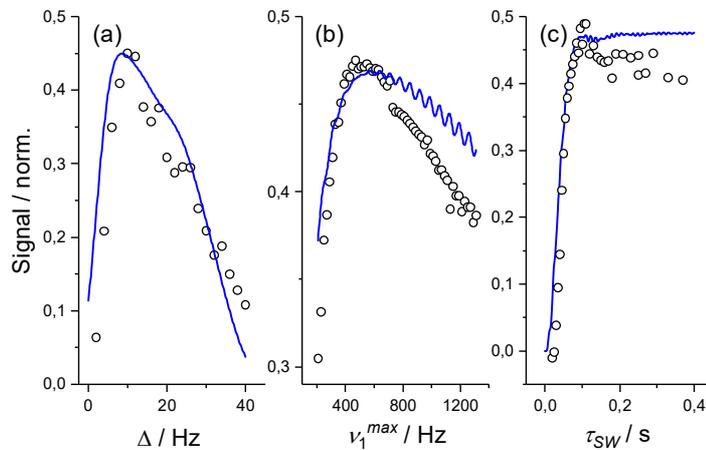

**Figure 13.** (top) Performance of APSOC upon variation of the RF-frequency $\Delta$ (A), its maximal amplitude $\nu_1^{max}$ (B), and switching time $\tau_{sw}$ (C). The optimal experimental parameters are $\Delta = 10$ Hz, $\nu_1^{max} = 410$ Hz, $\tau_{sw} = 95$ ms. All signals are normalized to the thermal NMR signal intensity.

## IV. Summary

In this work, a method for "fast" adiabatic switching of the spin Hamiltonian is presented, which is applied to optimize techniques for generating singlet spin order in NMR experiments. The idea behind optimization is to use "constant-adiabaticity" RF-ramps, which adjust to the properties of the spin Hamiltonian upon variation of an external parameter, which is the RF-field strength in the studied case. The method is tested on a specially designed molecule, which contains a pair of strongly coupled $^{13}$C spins having an extremely long lifetime of the singlet spin order. By using optimized RF-ramps we improved the performance of the APSOC and SLIC techniques; furthermore, the RF-ramps are robust to inaccuracies in setting the experimental parameters, such as the RF-strength, RF-frequency and switching time. The proposed optimization allowed us to approach the theoretically allowed limit of the M2S/S2M conversion efficiency, which is equal to 2/3. We anticipate that the proposed approaches are useful for singlet-state NMR and also for other experiments in magnetic resonance, which utilize adiabatic variation of the spin Hamiltonian, i.e., for the purpose of broadband excitation and for transferring hyperpolarized spin order from directly polarized spins to target spins.

## Conflicts of interest

There are no conflicts to declare.

## Acknowledgements

We acknowledge Prof. Malcolm H. Levitt and Dr. Giuseppe Pileio (University of Southampton) for stimulating discussions. This work has been supported by the Russian Foundation for Basic Research (grant No. 17-33-50077) and FASO of RF (project number 0333-2017-0002).

## Notes and references